# Prediction of the effective force on DNA in a nanopore based on density functional theory

Wen-Yue Tang, Guo-Hui Hu[*]

Shanghai Institute of Applied Mathematics and Mechanics, Shanghai Key Laboratory of Mechanics in Energy Engineering, Modern Mechanics Division, E-Institutes of Shanghai Universities, Shanghai University, 149 Yanchang Road, Shanghai 200072, People's Republic of China

ABSTRACT

We consider voltage-driving DNA translocation through a nanopore in the present study. By assuming the DNA is coaxial with the cylindrical nanopore, a hydrodynamic model for determining effective force on a single DNA molecule in a nanopore was presented, in which density functional theory (DFT) combined with the continuum Navier-Stokes (NS) equations is utilized to investigate electro-osmotic flow and the viscous drag force acting on the DNA inside a nanopore. Surface charge on the walls of the nanopore is also taken into account in our model. The consistence between our calculation and the previous experimental measurement indicates that the present theoretical model is an effective tool to predict the hydrodynamic resistance on DNA. Results show that charge inversion, which cannot be obtained by the Poisson-Boltzmann (PB) model, will reduce electro-osmotic velocity, or even lead to flow reversal for higher salt concentration. This is helpful to raise the effective force profoundly in the overscreening region.

Translocation of electrically driven single-stranded DNA and RNA molecules through an α–hemolysin channel of inside diameter of 2 nm was first demonstrated by Kasianowicz et al [1]. Each translocation through a nanopore can be detected by measuring variation of the ionic current caused by DNA blockage. This would provide a DNA sequencing method which is faster and cheaper than any previous one by many orders of magnitude. However, currently the translocation speed of individual polymer molecular is too fast for instruments to "read" each base signal: a single base pair transits the pore in about $\sim 10^{-8}$s under typical experimental conditions in solid-state nanopores. To better understand and overcome this difficulty, it is indispensable to predict the effective force applied on DNA inside and outside the nanopore.

In the DNA translocation process, the drag forces opposing electrical driving force can be divided into three components [2,3]: (1) the entropic force of DNA uncoiling/recoiling at the pore entrance/exits; (2) the hydrodynamic drag on untranslocated blob-like part of the DNA molecule outside the pore; (3) the electro-osmotic force acting on the linear DNA inside nanopore. Muthukumar et al [4] considered the entropic force arising from the conformational change of DNA molecule as the main factor determining the DNA translocation velocity. Storm et al. [5] evaluated the hydrodynamic drag acting on the blob-like DNA configuration outside the nanopore based on experimental parameters. His result yields a drag force of about 24pN when DNA's translocation velocity is 10mm/s for 16.5 $\mu m$ of double-stranded DNA. The magnitudes of these forces have been investigated by previous theories and experiments, which indicated that the forces (1) and (2) are smaller compared to the viscous drag inside the nanopore which offsets 80%-90% of the electrical driving force. Keyser et al. [6] experimentally measured the effective force acting on the DNA molecule inside nanopore in order to indirectly measure the viscous drag due to electro-osmotic flow, and obtained a value of 0.24±0.02pN mV$^{-1}$ for the force, which is independent of salt concentration from 0.02 to 1MKCl. This force corresponds to an effective charge of 0.50±0.05 electrons per base pair equivalent to a 70%-80% reduction of the total electrical driving force. A hydrodynamic model based on the Poisson-Boltzmann (PB) theory, proposed by Van Dorp et al. [7], was developed to calculate the effective force as a function of the nanopore radius. They found that although the theoretical curve trend based on the PB is qualitatively consistent with the experiments, it overestimates the stalling force by 50%. To get a quantitatively satisfied result, empirically reduction of surface potential difference was used to fit the experimental data in their study. Van den Hout et al. [8] argued, the failure of the PB modeling for the prediction of the effective force in the case of very small nanopores might be ascribed to that the finite ion size not considered in the model can't be ignored. Ghosal [9,10] developed a model for predicting translocation time of polyelectrolyte cross a nanopore. Electrophoretic speed of a polyelectrolyte is determined by a balance of electrical force and the viscous drag acting arising from electro-osmotic flow which calculated by PB equations. His results showed that the calculated translocation times are agreed with measurements in silicon nanopores.

---

[*] Email: ghhu@staff.shu.edu.cn, corresponding author.

Although the PB equation has been utilized in describing experimental results for monovalent counterion systems in the cylindrical nanopores with moderate surface charge densities, it breaks down when the surface electric potential and ionic concentration become larger. As an alternative approach, density functional theory (DFT) is an effective tool for investigating physical phenomena in the nanoscale under the continuous framework since it has the ability of containing detailed molecular interaction information once an appropriate intermolecular potential is given. DFT yields the time-mean density distributions of molecular species that minimize a free energy functional. DFT have been applicable for predicting the electric double layer (EDL) properties more accurate than the conventional PB theory due to incorporation of the ion size and electrical correlations. It has been successfully used to describe interesting phenomena that cannot be captured by the PB-based methods in EDL, such as layering and oscillations, charge inversion. The theory has been applied to EDLs in different geometries, such as planar [11], spherical [12], and cylindrical EDLs [13]. It has been reported that the results from DFT methods agree well with results by Monte Carlo and molecular dynamics simulation (MD) [12,14]. We aim at predicting the effective force by utilizing a theoretical model based on hydrodynamic equations and density functional theory in the present study.

Figure 1 shows the geometry of the present model. The parameters we considered are listed in the Table I [10]. We assumed that the nanopore is a cylindrical pore of identical diameter $R$ and length $L$=34 nm. The electric field intensity is obtained by assuming that the entire voltage drop of 120 mV occurs over the nanopore, thus, $E_0$=-3.53×10$^6$V/m. We model 16.5-μm-long dsDNA molecule as a negatively charged rigid cylinder, in which the average charge spacing on the DNA molecule is $b$ = 0.17 nm and the radius of DNA is $a$ = 1.0 nm. Our DFT method employs the molecular solvent model (MSM) [11] in which the ion species of the electrolyte are represented as centrally charged hard spheres, and the solvent is treated as a dense fluid of neutral hard spheres having a uniform dielectric constant $\varepsilon$=80. All three species have a molecular diameter of $d$ =

0.4nm. The corresponding value of the plasma constant, $\beta=e^2/4\pi k_B T \varepsilon \varepsilon_0 d$, is 1.785, where $k_B$ is the Boltzmann constant, $e$ is the magnitude of the elementary charge, $\varepsilon_0$ is the permittivity of free space, the temperature $T$ is 298 $K$. The Debye length is given by $\lambda_d=(4\pi\varepsilon\varepsilon_0 k_B T/e^2 \sum z_i C_{0i})^{-0.5}$. The dynamic viscosity of the electrolyte is taken as $\mu=1\times10^{-3}$Pa·s. The bulk molecular number density of the solvent is $\rho_3^0=0.7/d^3$. The salt concentration $C_0$ ranges from 0.02 to 4.0M. We assumed that DNA is coaxial with the cylindrical nanopore, and the flow field is uniform in the axial direction. We will calculate the effective force under a static condition, i.e. DNA translocation velocity $U_z=0$.

TABLE I. Physical parameters.

| Parameter | Symbol | Value |
|---|---|---|
| DNA length per electronic unit | b | 0.17nm |
| DNA diameter | a | 2.0nm |
| Pore surface charge density | σ | 0,-10,-20 mC/m² |
| Pore lengh | L | 34nm |
| Pore diameter | R | 1.75-17.0nm |
| Ion(solvent) diameter | d | 0.4nm |
| Solven dielectric constant | ε | 80 |
| Viscosity | μ | 1×10⁻³Pa·s |
| biased voltage | V | 120mV |

In classical DFT, the system reaches equilibrium $\overline{\rho_i}$ when the grand canonical potential $\Omega$ is at its minimum value. The equilibrium density distributions, $\rho_i(r)$, of multiple molecular species are determined by minimizing the grand potential energy $\Omega$,

$$\left.\frac{\delta\Omega[\{\rho_i\}]}{\delta\rho_i(r)}\right|_{\overline{\rho}}=0 \qquad (1)$$

In the present model, the grand potential is expressed as a functional of Helmholtz energy for density profile of certain species,

$$\Omega[\{\rho_i\}]=F^{id}[\{\rho_i\}]+F_{hs}^{ex}[\{\rho_i\}]+F_{ele}^{ex}[\{\rho_i\}]+F_C^{ex}[\{\rho_i\}]$$
$$+\sum_{i=1}^{N}d\mathbf{r}[V_{Di}(\mathbf{r})-\mu_i]\rho(\mathbf{r}) \qquad (2)$$

where $N$ is total number of particle species, $V_{Di}(\mathbf{r})$ is external field due to the DNA molecule. $\mu_i$ is the chemical potential of particle $i$. $F^{id}$ is the ideal gas contribution, $F_{hs}^{ex}$ account for the excess free energy due to hard sphere exclusions is derived from the modified fundamental measure theory [15], $F_{ele}^{ex}$ represents a coupling of Coulombic and hard-sphere interaction, $F_C^{ex}$ is the direct Coulomb contribution, given by

$$F_C^{ex}[\{\rho_i\}]=\frac{1}{2}\iint d\mathbf{r_1}d\mathbf{r_2}\sum_{i,j}\frac{z_i z_j e^2 \rho_i(\mathbf{r_1})\rho_j(\mathbf{r_2})}{\varepsilon\varepsilon_0|\mathbf{r_1}-\mathbf{r_2}|} \qquad (3)$$

where $z_i$ is the valence of the three-component system where including counterions, coions and solvent molecules, $z_i=\{1,-1,0\}$. The Euler-Lagrange equations is derives by using eq.(1) for the density profiles of particles [11-14]:

$$-k_BT \ln\left[\rho_i(\mathbf{r})/\rho_i^b\right] = \frac{\delta F_{hs}^{ex}}{\delta \rho_i(\mathbf{r})} - \mu_{i,hs}^{ex} + z_i e\left[\psi(r) - \psi^b\right]$$
$$-k_BT \sum_{j=1}^{N}\int d\mathbf{r}' \Delta C_{ij}^{(2)ele}\left(|\mathbf{r}-\mathbf{r}'|\right)\left(\rho_j(\mathbf{r}') - \rho_j^b\right), \quad (4)$$

Where $\{\rho_i^b\}$ is the bulk density of $i$, $\mu_{i,hs}^{ex}$ is excess chemical potential due to hard sphere correlation, $\Delta C_{ij}^{(2)ele}(r)$ is direct correlation function due to the residual electrostatic, which can be evaluated explicitly by the mean spherical approximate (MSA) [16,17]. $\psi(r)$ is the mean electrostatic potential (MEP) satisfying Poisson equation. For a cylindrical electric double layer, the MEP can be expressed in terms of the ionic density profiles:

$$\psi(r) = -\frac{4\pi e}{\varepsilon \varepsilon_0}\int_r^{R-\frac{d}{2}} dr' r' \ln\left(\frac{r'}{r}\right)\sum_i \rho_i(r') z_i \quad (5)$$

We focus on the steady fully developed electro-osmotic flows in the cylindrical pore once the MEP is given, the transverse velocity components $u_\theta$ and $u_y$ vanish, and $u_z$ become independent of axial position, which is written as [10],

$$u_z(r) = \frac{\varepsilon\varepsilon_0 E_0}{\mu}\left(\left(\psi - \psi|_{R-\frac{1}{2}d}\right) + \left(\psi|_{R-\frac{1}{2}d} - \psi|_{a+\frac{1}{2}d}\right)\left(\frac{\ln(r) - \ln(R-\frac{1}{2}d)}{\ln(a+\frac{1}{2}d) - \ln(R-\frac{1}{2}d)}\right)\right) \quad (6)$$

The velocity and the mean electrostatic potential are normalized in the following manner, $u^*=\mu e u_z/\varepsilon E_0 k_B T$, $\psi^*=-e\psi/k_B T$. $F_{driving}$ and $F_{drag}$ are the electric and viscous forces acting on the DNA inside the nanopore,

$$F_{driving} = eV/b \quad (7)$$

$$F_{drag} = 2\pi\mu L\left(a + \frac{1}{2}d\right)\frac{\partial u_z(r)}{\partial r}\bigg|_{a+\frac{1}{2}d} \quad (8)$$

From eq (7), the bare driving force $F_{driving}$ is about 112.9pN. The effective driving force can be obtained from $F_{eff} = F_{driving} - F_{drag}$.

The mean electrostatic potential (MEP) distribution computed by both DFT and PB model are presented in Fig.2*a* for various salt concentrations. We observed that at low salt concentration, the MEP reduce to zero monotonously along with radial direction, and the tendency of the velocity profiles from DFT model is similar to the prediction of the PB model, although the PB model overestimates the fluid velocity (as shown in Fig.2*b*). As salt concentration increasing, thickness of EDL becomes smaller, leading to the reduction of the electro-osmotic flow velocity. A significant phenomenon for higher salt concentration is occurrence of the charge inversion where the mean electrostatic potential change its sign near the DNA cylinder. Several DFT works [11,18] and MC or MD simulations [12,19,20] faithfully reproduces the phenomenon of charge inversion, which cannot be captured by alternative methods, such as the PB model or earlier version of DFT. The physical mechanism leading to charge inversion might be, due to strong excluded-volume effect or charge correlations, more than sufficient counterions stay near the cylindrical wall and overscreen the surface charge [18]. Thus the EDL can be divided into two layers. The first layer closed to the DNA surface is filled up with counterions. In the second layer, coions dominate the region due to their Coulomb attraction with the counterions in the first layer. Since the external electric forces applied to the ions in the two adjacent layers are opposite to each other, consequently it will weaken the resultant electro-osmotic flow. As the charge inversion become stronger with increasing salt concentration, flow reversal can even occur in the nanopore. Fig. 2*b* shows the velocity profile for $C_0$=2.0M becomes concave in the central, meanwhile negative local velocity can be observed in the curve for $C_0$=4.0M. To describe the process that ions neutralize the charges on DNA surface, a screening factor is defined as [20]

$$Q(r) = 2\pi b \int_0^r \sum_i z_i \rho_i(r') r' dr' \qquad (9)$$

$Q(r)$ represents the integral of the total charges over all species of ions within the annular volume extending radially from central axis to $r$ and axially over a length $b$ [21]. As shown in Fig. 2*c*, at low salt concentration, $Q(r)$ monotonously converges to unity according to electroneutrality condition. In the "overscreening" region, where the charge inversion occurs at high salt concentration, the screening factor overshoots the unity and reaches a maximum value $Q_{max}$, which essentially amounts to the strength of charge inversion.

In previous experiments and numerical simulations [7,8], researchers have discussed the net driving force for the different nanopore sizes. The nanopore transverse size determines the magnitude of the drag force acting on DNA, which can be attributed to a hydrodynamic coupling of the dsDNA counterions to the nanopore surface. Fig3*a* and *b* show the dependence of the effective force on pore radius at both low and high salt concentration (20mM and 1M) respectively. When the salt concentration is 1M, the effective force is found to decrease monotonically as the size of the nanopore was increasing. When DNA and the nanopore surface are separated enough, the effective driving force has a simple $[\ln((R-0.5d)/(a+0.5d))]^{-1}$ dependence on nanopore size based on eq (6), and the results are well consistent with experimental data [8]. When the salt concentration is 20mM, the Debye length is about $\lambda_d$=2.14 nm ($\lambda_d=(4\pi\varepsilon\varepsilon_0 k_B T/e^2\sum z_i C_{0i})^{-0.5}$) which comparable to the size of the nanopore. It is expected that the surface charge density of the nanopore would significantly influenced the results. Because of uncertainty of nanopore charge in the related experiment, we plot the effective force as a function of nanopore size with surface charge density of σ=-10mC/m$^2$ and -20mC/m$^2$ respectively. Fig.3*b* shows that our prediction captures the trends in the experimental data, and the results of σ=-20mC/m$^2$ looks more acceptable.

Previous theoretical analysis [3] based on the PB methods argues that the effective driving force decreases with the salt concentration. However, the phenomenon of charge inversion revealed by DFT model might have profound influence on the effective force in the overscreening region. To investigate this problem, we display the effective force per voltage as a function of the salt concentration for symmetric monovalent electrolytes based on the present DFT model (solid line) and PB model (dashed line) respectively. Meanwhile, $Q_{max}$ is used to parameterize the strength of charge inversion. Fig. 4*a* shows that for lower salt concentration, both DFT and PB models indicate that the effective forces decrease with salt concentration, due to the accumulation of counterions at the vicinity of the surface. Until $Q_{max}$ becomes larger than unity ($C_0$=1.0M for 1:1 salt), the effective force began to increase simultaneously as a consequence of the occurrence of charge inversion. Therefore, we conclude that the charge inversion has the effect of weakening the electro-osmotic flow, and reduces its hydrodynamic resistance, which consequently leads to the rise of the effective force. Although the result sounds physically reasonable, to our best knowledge, the corresponding phenomenon for high salt concentration had not been observed yet in the experiments available.

Uplinger et al. [22] observed adding $Mg^{+2}$ ions to 1.6 M KCl solution will lead to an increase in DNA translocation time, which implies the reduction of effective force. To investigate this phenomenon, we performed the effective force calculation for 2:1 salt electrolytes. As expected, $F_{eff}$ per voltage is generally smaller than the results for 1:1 salt electrolytes at lower salt concentration in Fig.4**a**, which is qualitatively agree with the results of Uplinger et al. [22]. The reason behind this phenomenon is that, the Debye layer becomes thinner in divalent ionic solution, which induces the increase of viscous resistance as a result. For higher concentration, however, it is easier for the divalent counterions to cause the charge inversion, and thus leads to the rise of effective force as the salt concentration increasing.

In conclusion**,** we have introduced a hydrodynamic model combined with density functional theory to study the effective driving force acting on a DNA molecule in a nanopore. The results obtained showed that the driving force decrease as the size of the nanopore increasing, which are quantitatively agree with the experiment available. We analyzed the dependence of the results on the pore diameter and salt concentration, as well as the ionic valence. An interesting finding is although the effective force decreases as the salt concentration if the concentration is lower, it will ascend in the oversrceening region. This phenomenon is ascribed to the charge inversion in the vicinity of DNA surface, which cannot be predicted by the conventional PB equation. The investigation of the influence of ionic valence indicates that in 2:1 salt electrolytes, the charge inversion and force rise occur at salt concentration smaller than that for 1:1 salt electrolytes. This study shows that, by considering the atomistic information, the continuum hydrodynamic model might provide an effective tool to examine the physical phenomenon in nanoscale.

Acknowledgments

This work was supported by the National Science Foundation of China (Grant No. 11272197), Research Fund for the Doctoral Program of Higher Education of China (Grant No. 20103108110004), and Shanghai Program for Innovative Research Team in Universities.

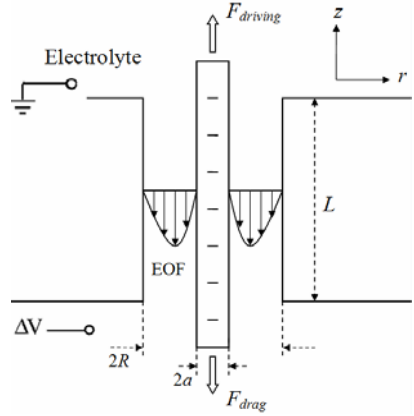

Fig.1. Cylindrical model of DNA in a nanopore

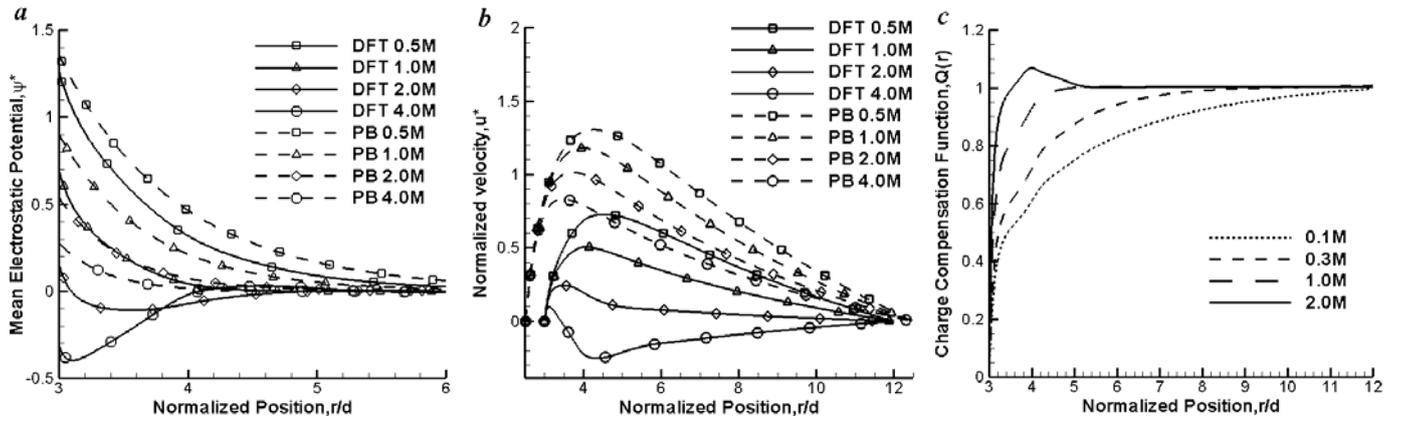

Figure.2. (*a*) The mean electrostatic potential distribution, (*b*) the electro-osmotic velocity field based on DFT (solid line) and PB model (dashed line). (*c*) Charge compensation function $Q(r)$ computed by DFT at various choices of 1:1 salt concentration with nanopore surface charge density 0 mC/m$^2$.

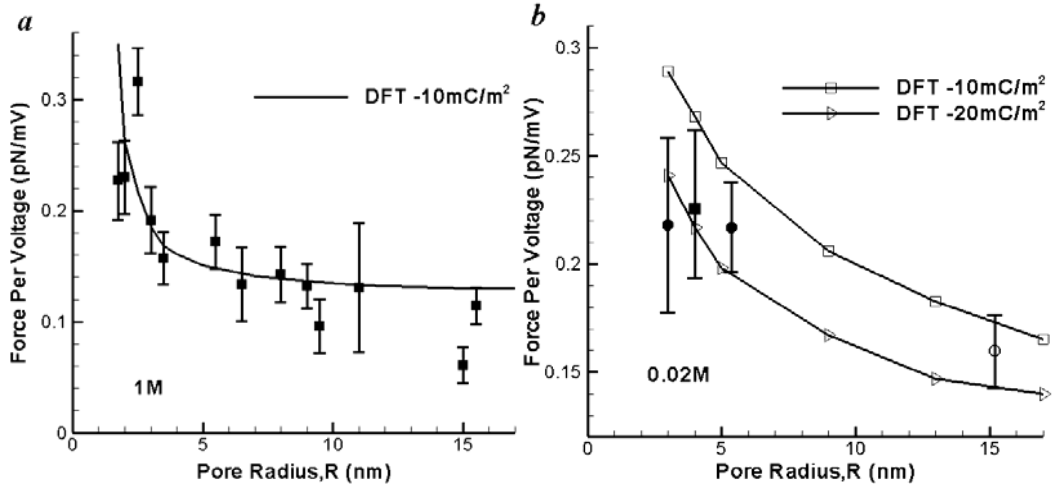

Figure.3. (*a*) The effective force per voltage versus pore radius. The solid lines represent the theoretical result for 1M salt concentration with surface charge density -10mC/m$^2$. (*b*) as same as (*a*) for 0.02M concentration with nanopore surface charge density -10,-20mC/m$^2$. The symbols correspond to the experiment data [7]: Squares: 20 mM KCl, hollow circles: 33 mM KCl, solid circles: 50 mM KCl.

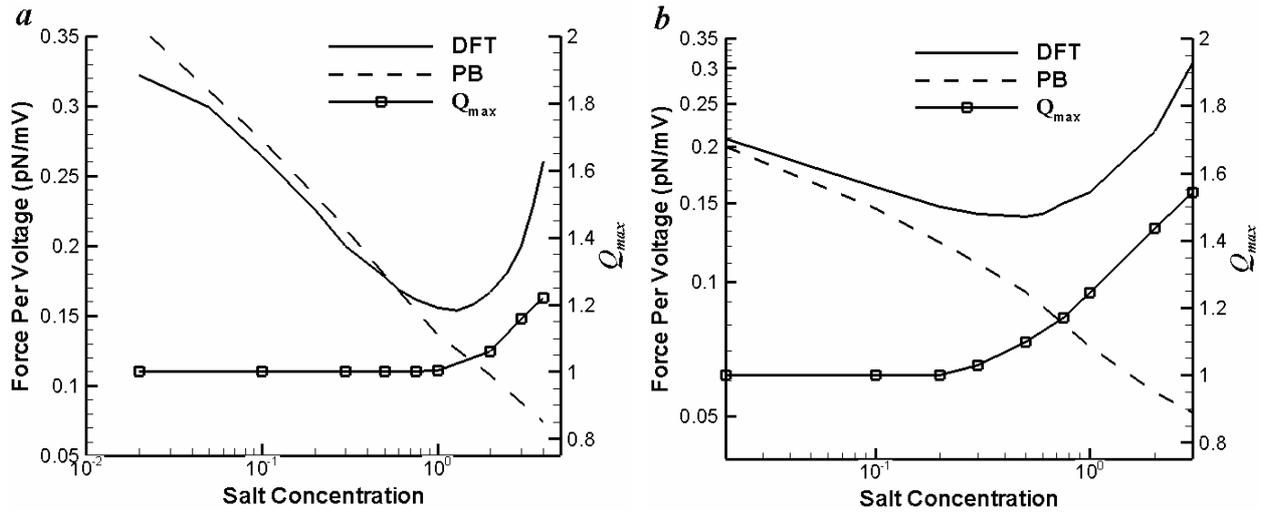

Figure.4. (*a*) The effective force per voltage versus salt concentration with varying nanopore surface charge density for 1:1 salt. (*b*) as same as (**a**) for 2:1 salt electrolytes.